\documentclass[prl,twocolumn,showpacs,superscriptaddress,amsfonts,amsmath,
floatfix]{revtex4}
\usepackage{graphicx}
\usepackage{psfrag}
\newcommand{\Journal}[4]{#1 {\bf #2}, #3 (#4)}
\newcommand{\PR}{Phys. Rev.}
\newcommand{\PRL}{Phys. Rev. Lett.}
\newcommand{\PRA}{Phys. Rev. A}

\newcommand{\JMP}{J. Math. Phys.}

\newcommand{\Science}{Science}

\begin{document}
\title {Motion of an impurity particle in an ultracold quasi-one-dimensional 
gas of hard-core bosons}
\author{M. D. Girardeau}
\email{girardeau@optics.arizona.edu}
\affiliation{College of Optical Sciences, University of Arizona, 
Tucson, AZ 85721, USA}
\author{A. Minguzzi}
\email{anna.minguzzi@grenoble.cnrs.fr}
\affiliation{Universit\'{e} Joseph Fourier, Laboratoire de Physique et 
Mod\'elisation des Mileux Condens\'es, C.N.R.S., B.P. 166, 38042 Grenoble, 
France}
\date{\today}
\begin{abstract}
The low-lying energy eigenstates of a one-dimensional (1D) system of many 
impenetrable point bosons 
and one moving impurity particle with repulsive zero-range impurity-boson 
interaction are found for all values of the impurity-boson mass ratio 
and coupling constant. The moving entity is
a polaron-like composite object consisting of the impurity clothed by a 
co-moving gray soliton. 
The special case with impurity-boson interaction of point hard-core 
form and impurity-boson mass ratio $m_i/m$ unity is first solved exactly as a 
special case of a previous Fermi-Bose (FB) mapping treatment of soluble 
1D Bose-Fermi mixture problems. Then a more general treatment
is given using
second quantization for the bosons and the second-quantized form of the FB 
mapping, eliminating the impurity degrees of
freedom by a Lee-Low-Pines canonical transformation. This yields the exact
ground state (total linear momentum $q=0$) and exact boson-impurity 
distribution function in the thermodynamic limit for arbitrary $m_i/m$ and 
arbitrary impurity-boson interaction strength. These results are then
extended to states with $q>0$.
\end{abstract}
\maketitle
Due to the rapidly increasing sophistication of experimental techniques
for probing ultracold gases, theoretical emphasis has shifted from effective 
field approaches 
to more refined methods capable of dealing with correlations. 
When confined in a de Broglie waveguide with transverse trapping so tight that
longitudinal energies are less than the transverse vibrational excitation 
energy quantum $\hbar\omega_\perp$, an ultracold Bose gas 
becomes effectively one-dimensional (1D) with a confinement-induced resonance 
in the 1D scattering length \cite{Ols98}. 
In units such that $\hbar=1$, the dimensionless 1D coupling constant 
is $\gamma_B=mg_{1D}^B/n$ where $g_{1D}^B$ is the coupling constant for 
zero-range interactions $g_{1D}^B\delta(x_j-x_\ell)$ \cite{Ols98} of 
Lieb-Liniger (LL) form \cite{LieLin63}, and $n$ is the 1D density $n=N/L$ with
$N$ the number of bosons and $L$ the cell length for periodic boundary 
conditions. At low densities where $\gamma_B\gg 1$, the boson-boson 
interaction reduces to impenetrable point form,
the so-called Tonks-Girardeau (TG) limit for which the exact many-body ground
and excited states were found some 48 years ago by the Fermi-Bose (FB)
mapping method \cite{Gir60Gir65}. The fermionization predicted therein has
recently been experimentally confirmed \cite{Par04Kin04,Kin0506}, leading
to much recent theoretical and experimental activity on various properties of 
such strongly-correlated quasi-1D ultracold gases. It was found recently 
\cite{Chi} that impurity atoms in
an ultracold Bose gas with velocity $v_i$ have collision cross sections 
with the bosons which increase dramatically when $v_i>c$ where $c$ is the
speed of sound in the Bose gas. Here we 
shall investigate the behavior of a moving impurity particle
in an ultracold 1D gas of impenetrable point bosons, assuming a repulsive,
zero-range impurity-boson interaction of arbitrary strength, and arbitrary
impurity-boson mass ratio. 

{\it Hard-core impurity in a TG gas:} Consider first the
case $N_F=1$ of our previous exact solution of the problem of $N$ hard-core 
bosons and $N_F$ hard-core fermions in 1D \cite{GirMin07}. 
One starts from a ``model wavefunction'' 
$\Psi_M=\sum_P \varepsilon(P)u_1(Px_1)\cdots u_{N}(Px_{N})u_{N+1}(y)$ where
$y$ is the impurity position and 
the sum runs over all $(N+1)!$ possible permutations of these variables 
\emph{including permutations exchanging bosons with the impurity},  
$\varepsilon(P)$ is the usual $\pm 1$ sign of the permutation, and
$u_1,\cdots,u_{N+1}$ are $N+1$ orthonormal orbitals occupied by all $N$ bosons 
plus the one impurity. $\Psi_M$ vanishes at all the 
points $x_j=x_\ell$ and $x_j=y$ required by the hard-core constraints,
and its improper symmetry under boson-boson and boson-impurity 
exchange can be repaired by a generalized FB mapping function 
$A=\prod_{1\le j<\ell\le N}\text{sgn}(x_j-x_\ell)
\prod_{j=1}^N\text{sgn}(x_j-y)$ \cite{GirMin07}
where the sign function $\text{sgn}(x)$ is $+1\ (-1)$ if $x>0\ (x<0)$. 
Then the physical wavefunction is 
$\Psi(x_1,\cdots,x_N;y)=A(x_1,\cdots,x_N;y)\Psi_M(x_1,\cdots,x_N;y)$. 
The model wave function 
$\Psi_M$ is an exact many-body 
energy eigenstate if the orbitals $u_{\nu}(x)$ are eigenfuctions of the
single-free-particle Hamiltonian 
$-\frac{1}{2m}\frac{\partial^2}{\partial x^2}$ with eigenvalues
$\epsilon_{\nu}$, and  the physical state $\Psi=A\Psi_M$ is an
energy eigenstate with eigenvalue $\sum_{\nu}\epsilon_{\nu}$. 
If there is no
external potential and the system is on a ring of circumference $L$ with the
$x_j$ and $y$ measured circumferentially, 
the $u_\nu$ are plane waves $L^{-1/2}e^{ik_\nu x}$
with $k_\nu$ which are integer (+,-,0) multiples of $\frac{2\pi}{L}$
and $\epsilon_\nu=\frac{k_\nu^2}{2m}$.

\begin{figure}
  \centering
\psfrag{P/2mc}{$q/2mc$}
\psfrag{E/(mc^2/2)}{$E/(mc^2/2)$}
\includegraphics[width=7.5cm,angle=0]{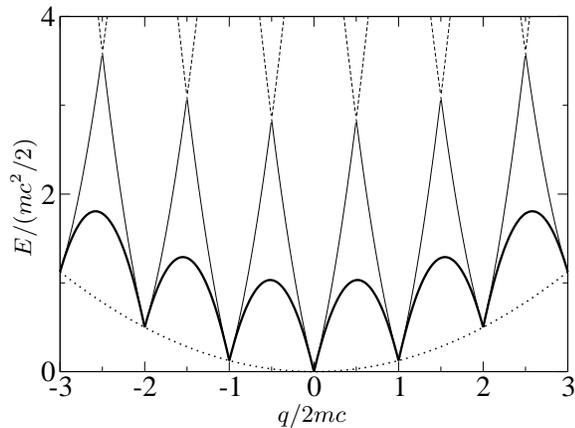}
  \caption{Excitation energy branches as functions of the momentum transfer
  $q$ for $N+1=31$ particles. The bold solid lines indicate the
  lowest-energy (hole) excitations, the thinner solid 
  lines indicate the lowest excitations of both Fermi-surface (FS) and 
  umklapp types, the dashed lines are the prolongations of the FS and 
  umklapp excitations to higher-energy values, and the underlying dashed
  parabola is $q^2/2(N+1)m$.
}
  \label{fig1}
\end{figure}

Assuming $N$ to be even, the model ground state $\Psi_{0M}$ is a filled Fermi
sea of the $N+1$ lowest plane-wave orbitals with $-k_F\le k_\nu\le k_F$
where $k_F=n\pi$ and $n=N/L$, i.e.,
$\nu=-\frac{N\pi}{L},-\frac{(N-2)\pi}{L},\cdots,\frac{(N-2)\pi}{L},
\frac{N\pi}{L}$ and has total linear (or angular, on a ring) momentum zero.
For this moving impurity problem, one needs excited states of
nonzero total momentum $q$. The energy landscape is identical with that of
$N+1$ TG bosons; only the physical interpretation is different. These
excitations are the $\gamma_B\to+\infty$ limit of the types I and II
excitations of the LL gas, Fig. 4, p. 1620 of \cite{LieLin63}, and are shown 
in Fig.\ref{fig1}. 
The lowest excitations, indicated by the bold solid lines,
correspond to the type II excitations of \cite{LieLin63} and 
correspond to promoting a particle from $k_F-q+2\pi/L$
to $k_F+2\pi/L$ or from $-k_F+|q|-2\pi/L$ to $-k_F-2\pi/L$, thus leaving
a hole in the interior of the Fermi sea. 
As $q$ increases from zero
to $2k_F$ the hole moves to the left from $k_F$ to $-k_F$, at which point
the whole Fermi sea has been shifted to the right by $2\pi/L$.
Here $c=nh/2m$ is the speed of sound in the TG gas \cite{Gir60Gir65}.
The excitation energy is periodic in the center of mass system with 
period $2mc=2k_F$ \cite{Blo73}, and the solid lines for $|q|>2mc$ correspond to
this
periodicity, the underlying dashed parabola $q^2/2(N+1)m$ being the result
of transformation from the center of mass to the laboratory system, and
the bold solid lines correspond to hole excitations from the original and
displaced Fermi surfaces $k_F+j2\pi/L$ and $-k_F+j2\pi/L$ with 
$j=0,\pm 1,\pm 2,\cdots$. When $j$ reaches $(N+1)/2$ the flow speed
$v_j=j2\pi/mL$ reaches the speed of sound $c$, the slope in the entire interval
$q_{j-1}<q<q_j$ becomes
$\ge 0$ so that the energy barrier disappears, metastability is lost, and Mach 
wave drag commences. The boson-impurity distribution function
$\rho_{bi}(x-y)=N(N-1)\int|\Psi(x,x_2,\cdots,x_N;y)|^2dx_2\cdots dx_N$
is found in the thermodynamic limit to be
$\rho_{bi}(x-y)=n^2[1-j_0^2(k_f(x-y))]$ where $j_0(\xi)=\sin\xi/\xi$ is the
spherical Bessel function of order zero. This is independent of the flow speed
$v_j$ and agrees with \cite{Gir60Gir65}.

{\it Quantized field representation:} Generalize now to the case where 
impurity and boson masses are unequal and boson-impurity interactions are 
of the form $\lambda_{bi}\delta(x_j-y)$. The impenetrable point boson-boson
interaction will eventually be used, but initially
assume it to be $\lambda_b\delta(x_j-x_\ell)$. The method used is motivated by 
a previous 3D weak-coupling treatment 
\cite{Gir61}. We use second quantization for
the bosons while retaining Schr\"{o}dinger representation for the impurity, 
with position variable $y$ and momentum operator 
$\hat{p}_i=\frac{1}{i}\frac{\partial}{\partial y}$. Then the Hamiltonian is
\begin{eqnarray}\label{2ndQB}
&&\hat{H}_B=\int dx\ \hat{\psi}_B^\dagger(x)
\left[-\frac{1}{2}\frac{\partial^2}{\partial x^2}\right]
\hat{\psi}_B(x)\nonumber\\
&+&\frac{\lambda_b}{2}\int dx\ [\hat{\psi}_B^\dagger(x)]^2\hat{\psi}_B^2(x)
-\frac{1}{2m_i}\frac{\partial^2}{\partial y^2}
+\lambda_{bi}\hat{\rho}_B(y)
\end{eqnarray}
where $\hat{\psi}_B(x)$ and $\hat{\psi}_B^\dagger(x)$ are the Bose field
annihilation and creation operators, 
$\hat{\rho}_B(y)=\hat{\psi}_B^\dagger(y)\hat{\psi}_B(y)$ is the boson density 
operator at the impurity position, and we assume units wherein the boson mass 
$m=1$. 
Now enlarge the state space to also contain Fock states with Fermi
field creation operators $\hat{\psi}_F^\dagger(x)$ where $\hat{\psi}_F$
and $\hat{\psi}_F^\dagger$
satisfy the usual anticommutation relations, but commute with the Bose
fields, and both Bose and Fermi fields commute with the impurity position
and momentum operators. On this enlarged Hilbert space, introduce a canonical
transformation which is the second-quantized form \cite{GanPus08} of the
FB mapping \cite{Gir60Gir65}:
\begin{equation}
\hat{\psi}_B(x)=e^{i\pi\int_{-\infty}^x dx'\hat{\rho}_B(x')}\hat{\psi}_F(x) 
\end{equation}
Then $\hat{H}_B$ is transformed into
\begin{equation}\label{2ndQF}
\hat{H}_F=\int dx\ \hat{\psi}_F^\dagger(x)
\left[-\frac{1}{2}\frac{\partial^2}{\partial x^2}\right]
\hat{\psi}_F(x)
-\frac{1}{2m_i}\frac{\partial^2}{\partial y^2}
+\lambda_{bi}\hat{\rho}_F(y)
\end{equation}
where $\hat{\rho}_F(y)=\hat{\psi}_F^\dagger(y)\hat{\psi}_F(y)$ is the fermion 
density operator at the impurity position. The zero-range boson-boson
interaction term in $\hat{H}_B$ has disappeared because its transform
involves $\hat{\psi}_F^2(x)$ and its hermitian conjugate, which vanish by
the anticommutation relations. Because of nonlocality of the FB mapping
transformation, the transform of the boson kinetic energy operator does not
have the simple form above on the whole
Hilbert space, but on the subspace of states $|\Phi_B\rangle$ satisfying the 
impenetrable point hard core
constraint $\hat{\psi}_B^2(x)|\Phi_B\rangle=0$ it reduces to this 
form \cite{Gir60Gir65,GanPus08}. 

Next we make a second canonical transformation of the type used by Lee, Low, 
and Pines (LLP) in their theory of the polaron \cite{LLP53Gir83} which takes 
advantage of total linear momentum 
conservation so as to eliminate the impurity dynamical variables, as in
\cite{Gir61}. It is effected by a unitary operator 
$\hat{U}=e^{-iy\hat{p}_F}$ where 
$\hat{p}_F=\int dx\ \hat{\psi}_F^\dagger(x)
\frac{1}{i}\frac{\partial}{\partial x}\hat{\psi}_F(x)$ is the
fermion momentum operator, with the result 
\begin{equation}\label{LLP}
\hat{U}^{-1}y\hat{U}=y\ ,\ \hat{U}^{-1}\hat{p}_i\hat{U}
=\hat{p}_i-\hat{p}_F
\ ,\ \hat{U}^{-1}\hat{\psi}_F(x)\hat{U}=\hat{\psi}_F(x-y)\ .
\end{equation}
Then (\ref{2ndQF}) is transformed into 
\begin{equation}\label{scriptH}
\hat{\mathcal{H}}=\hat{U}^{-1}\hat{H}_F\hat{U}=\hat{T}_F
+\lambda_{bi}\hat{\rho}_F(0)
+\frac{(\hat{p}_i-\hat{p}_F)^2}{2m_i}
\end{equation}
where 
$\hat{T}_F$, the fermion kinetic energy operator, is the first term of
Eq. (\ref{2ndQF}). 
Since $y$ is absent from $\hat{\mathcal{H}}$, $\hat{p}_i$ commutes with
$\hat{\mathcal{H}}$ and may be replaced by its eigenvalue $q$. 
Any energy eigenstate of $\hat{\mathcal{H}}$ with eigenvalue $E$ 
can be written as a direct product
$|\Phi\rangle=L^{-1/2}e^{iqy}|\Phi_F\rangle$ where $|\Phi_F\rangle$ is 
independent of $y$. Then defining $|\Phi_q\rangle=\hat{U}|\Phi\rangle$ one 
finds that $\hat{H}_F|\Phi_q\rangle=E_q|\Phi_q\rangle$ and 
$\hat{P}|\Phi_q\rangle=q|\Phi_q\rangle$ provided that
$\hat{\mathcal{H}}_q|\Phi_q\rangle=E_q|\Phi_q\rangle$, where
$E_q$ is the energy eigenvalue of the given direct product eigenstate of
$\hat{\mathcal{H}}$, 
$\hat{P}=\hat{p}_i+\hat{p}_F$ is the conserved \emph{total} linear momentum 
(impurity plus fermions), and $\hat{\mathcal{H}}_q$ is the Hamiltonian
obtained from $\hat{\mathcal{H}}$ by replacing the impurity momentum operator
$\hat{p}_i$ by the c-number $q$. With this replacement 
Eq. (\ref{scriptH}) becomes 
\begin{equation}\label{H_q}
\hat{\mathcal{H}}_q=\hat{T}_F+\lambda_{bi}\hat{\rho}_F(0)
+\frac{q^2}{2m_i}-\frac{q\hat{p}_F}{m_i}+\frac{\hat{p}_F^2}{2m_i}\ .
\end{equation}
After mapping back to the physical Hilbert space (bosons plus impurity), $q$ 
is found to be the \emph{total} linear momentum in the laboratory frame.
The ground state has $q=0$, in which case two of the last three terms in
(\ref{H_q}) vanish identically. We will consider that case now and 
will generalize later to the case $q>0$. 
For $q=0$ the mean fermion momentum $\langle\hat{p}_F\rangle$ vanishes by
symmetry, so the expectation value of the last term in (\ref{H_q}) is 
proportional to the mean square fluctuation of $\hat{p}_F$.
This fluctuation term 
will be shown later to be negligible in the thermodynamic limit, and will
be dropped now.

{\it Orbitals:} With the last two terms dropped and the zero of energy shifted 
by $q^2/2m_i$, Eq. (\ref{H_q}) is the Hamiltonian of free
fermions in the field of a potential $\lambda_{bi}\delta(x)$ centered on the 
impurity. Its $N$-particle ground state is a filled Fermi 
sea of the lowest $N$ orbitals $\phi_k(x)$ of the single-particle
Schr\"{o}dinger equation $\hat{H}\phi_k(x)=\epsilon_k\phi_k(x)$ with  
\begin{equation}\label{Sch}
\hat{H}=-\frac{1}{2}\frac{\partial^2}{\partial x^2}
+\lambda_{bi}\delta_{\text{per}}(x)
\end{equation}
where $\delta_{\text{per}}$ is an $L$-periodicized delta function,
$\delta_{\text{per}}(x)=\sum_{\nu=-\infty}^\infty\delta(x+\nu L)$.
Our approach here is a generalization
of a previous treatment of dark solitons in a TG gas \cite{GirWri00b}.
As in \cite{Gir60Gir65} we assume that $N$ is odd so that the mapping back
to the boson Hilbert space will be periodic. 
It is convenient to choose real eigenstates of (\ref{Sch}) since the
$N$-fermion ground state has zero total linear momentum, and this will be
satisfied automatically if all the orbitals have zero flux.  
We generalize Eq. (6) of \cite{GirWri00b}
to
\begin{eqnarray}\label{orbitals}
\phi_{k}^{(+)}(x) & = & \mathcal{N}_k[\sin(k|x|)+A_k\cos(kx)]  , \nonumber\\
\phi_{k}^{(-)}(x) & = & \sqrt{2/L}\sin(kx)
\end{eqnarray}
where the allowed $k$-values for both the even-parity orbitals $\phi_{k}^{(+)}$
and the odd-parity orbitals $\phi_{k}^{(-)}$
are $k=2\pi/L,4\pi/L,\cdots$. These orbitals satisfy periodic boundary
conditions on the interval $-L<x<L$, and are to be
periodically extended to all other periodicity cells \cite{Note1}.
The delta function potential induces cusps in the even orbitals 
$\phi_{k}^{(+)}$ at $x=0$ and more generally
$x$ which are integral multiples of $L$, which are to be determined from 
$\phi_k\ '(0+)-\phi_k\ '(0-)=2\phi_k\ '(0+)=2\lambda_{bi}\phi_k(0)$ 
where the prime denotes the derivative, thus determining $A_k$ and hence the 
normalization constant $A_k$. One finds $A_k=k/\lambda_{bi}$ and 
$\mathcal{N}_k=\sqrt{2/L(1+A_k^2)}=\lambda_{bi}\sqrt{2/L(\lambda_{bi}^2+k^2)}$.
As in \cite{GirWri00b}, the odd orbitals $\phi_{k}^{(-)}$ do not see the
delta potential and are cusp-free.

\begin{figure}
\centering
\psfrag{rhobixyovern}{{\large $\rho_{bi}(x-y)/n$}}
\psfrag{nxminusy}{{\large $(x-y)\ n$}}
\includegraphics[width=7.5cm,angle=0]{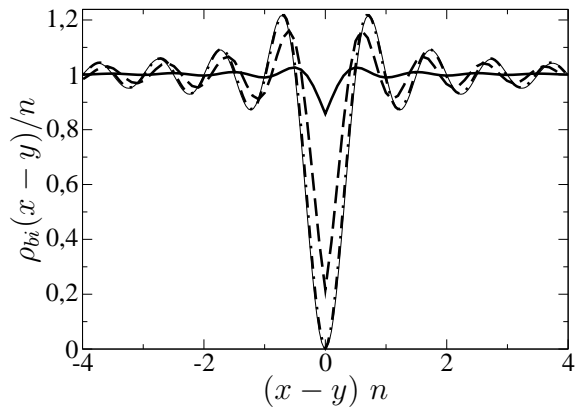}
  \caption{Impurity-boson distribution function for
various choices of impurity-boson coupling constant $\tilde \lambda_{bi}$.
Thin solid line: $\tilde \lambda_{bi}= \infty$ from analytical expression.
Other curves from numerical calculations with $\frac{N+1}{2}=200$ orbitals:
thick solid line $\tilde \lambda_{bi}=0.1$, 
dashed line $\tilde \lambda_{bi}=1$, dot-dashed line $\tilde \lambda_{bi}=10$.}
  \label{fig2}
\end{figure}

{\it Impurity-boson distribution function:} 
By Eq. (\ref{LLP}), the impurity-boson distribution function $\rho_{bi}$
in the laboratory system is the expectation value of
 ${\hat{\psi}}_F^\dagger(x-y){\hat{\psi}}_F(x-y)$ where $y$ is the impurity
position. Then 
$\rho_{bi}(x-y)=\sum_{\text{occ}}|\phi_k(x-y)|^2$ where 
$\sum_{\text{occ}}$ runs over the $N$ occupied orbitals with 
$\frac{2\pi}{L}\le k\le (N+1)\pi/L$. 
By Eq. (\ref{LLP}) and the
argument following Eq. (\ref{scriptH}), the 
wave function in the laboratory system is
$\Phi_{\text{lab}}(x_1,\cdots,x_N;y)=L^{-1/2}e^{iqy}
\Phi_F(x_1-y,\cdots,x_N-y)$ where $\Phi_F$ is the filled 
Fermi sea. All of the linear momentum resides
in the prefactor $e^{iqy}$ since the factor $\Phi_F$ is now translationally
invariant. Application of the standard definition
$\rho_{bi}(x-y)=N(N-1)\int|\Phi_{\text{lab}}(x,x_2,\cdots,x_N;y)|^2
dx_2\cdots dx_N$ verifies the correctness of the above expression
for $\rho_{bi}$ as a sum over density contributions from all occupied 
orbitals. In the thermodynamic limit the sum over the trivial orbitals
with $A_k=0$
is an integral over the Fermi sea from $0$ to $k_F$, yielding 
$\frac{n}{2}[1-j_0(2k_F(x-y))]$ where $j_0(z)=\sin(z)/z$ and 
$n=\frac{N}{L}=\frac{k_F}{\pi}$
is the mean fermion number density. In the limiting cases 
${\tilde{\lambda}}_{bi}=\lambda_{bi}/k_F=0$ 
and ${\tilde{\lambda}}_{bi}\to\infty$ the
sum over the orbitals with $A_k\ne 0$ can also be evaluated analytically
in the thermodynamic limit; combining with the above contribution
one finds $\rho_{bi}(x-y)=k_F/\pi=n$ for ${\tilde{\lambda}}_{bi}=0$
as expected, and for ${\tilde{\lambda}}_{bi}\to\infty$ (impenetrable impurity)
one finds $\rho_{bi}(x-y)=n[1-j_0(2k_F x)]$ which vanishes when $x-y=0$ as
expected. 
For intermediate values of ${\tilde{\lambda}}_{bi}$ the
sum over the $\frac{N+1}{2}=200$ lowest orbitals with $A_k\ne 0$ has been 
evaluated numerically, and 
adding the above expression for the $A_k=0$ contribution yields the results
shown in Fig. \ref{fig2} for $\rho_{bi}(x-y)$.
These results differs from the first section (one-fermion case of
the boson-impurity mixture theory \cite{GirMin07}) because the many-body
wave function there represented the impurity ``dissolved'' into the
TG gas with the total linear momentum $q$ shared equally by the impurity
and $N$ bosons, whereas here the impurity moves through a nonmoving sea of 
fermions.

{\it Generalization to $q>0$}: Here we  
generalize to the case of a moving impurity by including the
term $-\frac{q\hat{p}_F}{m_i}$ in Eq. (\ref{H_q}).
Dropping $\hat{p}_F^2/2m_i$
and the c-number
$\frac{q^2}{2m_i}$, one finds that the $N$-particle ground state 
is a filled Fermi 
sea of the lowest $N$ orbitals $\phi_k(x)$ of the single-particle
Schr\"{o}dinger equation $\hat{H}_q\phi_k(x)=(k^2/2)\phi_k(x)$ with  
\begin{equation}\label{genSch2}
\hat{H}_q=
\frac{1}{2}\left(\frac{1}{i}\frac{\partial}{\partial x}
-Q\right)^2+\lambda_{bi}\delta_{\text{per}}(x)\ 
\end{equation}
where 
$Q=q/m_i$. 
The momentum shift $Q$ can be eliminated by a gauge 
transformation, writing $\phi_k(x)=e^{iQx}u_k(x)$, after which $u_k$
is an eigenfunction of the original Schr\"{o}dinger equation (\ref{Sch}). 
However, now periodicity of $\phi_k$ is not automatic, but must be
enforced, leading to nontrivial allowed values of $k$ determined by a
secular equation. This 
greatly complicates the solution. However, simple results can be obtained
in the thermodynamic limit by noting that the total momentum $q$ is
necessarily an integral multiple of $2\pi/L$, i.e., $q=\nu 2\pi/L$ with $\nu$
a non-negative integer. For finite, nonzero $q$, the thermodynamic limit is
then to be attained by letting the integer $\nu\to\infty$ and fermion number
$N\to\infty$ in such a way that $\nu/N\to q/2\pi n$ where $n=N/L$ is 
the fermion density in the thermodynamic limit where $N\to\infty$, 
$L\to\infty$, $N/L\to n$. In that limit there are infinitely many values of
$m_i$ which are divisors of $\nu=Lq/2\pi$, and the
set of all such divisors is dense on the positive real axis. For all such
$m_i$, $Q$ is an integral multiple of $2\pi/L$, the orbitals reduce
to the $q=0$ orbitals (\ref{orbitals}), the mean fermion momentum
$\langle\hat{p}_F\rangle =0$, and the impurity-boson distribution function
reduces to that of Fig. \ref{fig2}.

{\it Momentum fluctuation and thermodynamic limit}:  
Here we show that the term $\frac{\hat{p}_F^2}{2m_i}$ in (\ref{H_q}) 
is negligible. $\langle\hat{p}_F\rangle$ is zero since the orbitals 
(\ref{orbitals}) are real and have zero flux; hence 
$\langle\hat{p}_F^2\rangle=\langle\Psi_0|\hat{p}_F^2|\Psi_0\rangle$ 
is the mean square fluctuation of the 
fermion momentum about its mean. The $N$-fermion ground state is
$|\Psi_0\rangle=(\prod_{k}\hat{c}_{k}^\dagger)|0\rangle$ where the 
$\hat{c}_{k}^\dagger$ are creation operators
for the $N$ lowest orbitals $\phi_{k}$ of Eq. (\ref{orbitals}).
One has 
$\hat{p}_F=\sum_{kk'}
\hat{c}_k^\dagger(\phi_k|\frac{1}{i}\frac{\partial}{\partial x}|\phi_{k'})
\hat{c}_{k'}$. The diagonal elements 
$(\phi_{k}^{(+)}|\frac{1}{i}\frac{\partial}{\partial x}|\phi_{k}^{(+)})$ and 
$(\phi_{k}^{(-)}|\frac{1}{i}\frac{\partial}{\partial x}|\phi_{k}^{(-)})$
vanish since their integrands are odd, and the off-diagonal elements lead to
$\hat{p}_F
=\sum_k[\frac{-ikA_k}{2\sqrt{1+A_k^2}}\hat{c}_k^{(+)\dagger}\hat{c}_k^{(-)}
+\text{h.c.}]+\sum_{kk'}[\mathcal{O}(L^{-1})
\hat{c}_k^{(+)\dagger}\hat{c}_{k'}^{(-)}+\text{h.c.}]$ where 
$\mathcal{O}(L^{-1})$ is a function of $k$ and $k'$ which is proportional
to $L^{-1}$ in the thermodynamic limit where $N\to\infty$ and $L\to\infty$
with $n=N/L$ fixed. The $\sum_k$ vanishes when acting on 
$|\Psi_0\rangle$, since both $\phi_{k}^{(+)}$ and $\phi_{k}^{(-)}$ are 
occupied below the Fermi surface and both empty above it. 
There are two factors of $L^{-1}$ from the other term and two
factors of $L/2\pi$ from $\sum_{kk'}\to (L/2\pi)^2\int dkdk'$, so 
$\frac{1}{2m_i}\langle\Psi_0|\hat{p}_F^2|\Psi_0\rangle=\mathcal{O}(1)$ in the
thermodynamic limit, negligible compared with the total ground 
state energy which is $\mathcal{O}(L)$. We expect that the 
contribution of this term to $\rho_{bi}(x-y)$ is only $\mathcal{O}(L^{-1})$.
The mean square fluctuation of the impurity momentum $\hat{p}_i$ is also 
$\mathcal{O}(1)$ since the total momentum $q$ is conserved.

{\it Prospects}: We have obtained exact results in the thermodynamic limit
for the ground state energy and impurity-boson distribution function
of an impurity in a TG gas, for arbitrary impurity-boson mass ratio and
interaction strength. Control of these two additional parameters
should provide an arena for future experiments building on the pathbreaking
experiments \cite{Par04Kin04,Kin0506}.
\begin{acknowledgments}
The first part of this paper is an outgrowth of work \cite{GirMin07}
initiated when we were participants in the 2007 workshop ``Quantum Gases'' at 
the Institut Henri Poincar\'{e}-Centre Emile Borel (IHP) in  Paris, and
we are grateful to the workshop organizers and the IHP for hospitality and 
support. We also thank Ewan Wright for helpful comments. Research of A.M.
is supported by the Centre National de la Recherche Scientifique (CNRS).
\end{acknowledgments}
\end{document}